# Moyal Quantization, Holography, and the Quantum Geometry of Surfaces

George Chapline[+] and Alex Granik[++]

Abstract

An elementary introduction is provided to the phase space quantization method of Moyal and Wigner. We generalize the method so that it applies to 2-dimensional surfaces, where it has an interesting connection with quantum holography. In the case of Riemann surfaces the connection between Moyal quantization and holography provides new insights into the Torelli theorem and the quantization of non-linear integrable models. Quantum holography may also serve as a model for a quantum theory of membranes.

**1. Introduction**

It has been known for some time that there is a close connection between ordinary quantum mechanics, $W(\infty)$ transformations, and the geometric quantization of non-linear integrable models [1, 2, 3, 4]. This paper is intended as an elementary introduction to these connections. In section 2 we provide an introduction to the phase space quantization method of Moyal [5] and Wigner [6] and its connection with area preserving transformations of a surface. In particular we show that Moyal - Wigner quantization can be viewed as the unique form of Born-Jordan quantum mechanics that respects the area preserving feature of Hamiltonian flows. We also show that, in the case of a harmonic oscillator, Moyal -Wigner quantization provides a simple way to transform from the usual p and q representation to action - angle variables. In section 3 we discuss the problem of quantizing the parameterization of a curved smooth two dimensional surface.

This problem is most interesting in the case of a Riemann surface with non-trivial topology. Such surfaces can also be represented holographically, and indeed holographic representations of a Riemann surface can be thought of as a pedestrian version of the mathematically elegant characterization a Riemann surface in terms of integrals of harmonic differentials and Jacobian varieties [7]. In section 3 we show that in a paraxial ray type approximation for the electromagnetic field the usual mode variables for the electromagnetic field can be replaced by field variables E and $E^+$ whose transverse variation represents the structure of a hologram. These field variables depend on both the position on the hologram and the orientation of the illuminating laser beam, and quantum fluctuations of these variables reflect quantum "fuzziness" in the surface. This may perhaps be interpreted by saying that the classical geometry of the Riemann surface has been replaced with quantum geometry. However, the classical topology of the surface is preserved even in the quantum theory, and this provides a remarkable connection between phase space quantization and the geometric quantization of non-linear integrable systems.

**2. Moyal Quantization and Area-Preserving Transformation.**

The problem of transformation of classical functions defined on the phase space p-q into the respective quantum operators had been addressed at the very dawn of quantum mechanics. In 1925 M.Born and P.Jordan [8] introduced what looks like the first quantization method applied to the classical expression $f(p,q) = p^m q^n$:

$$p^m q^n \to \mathbf{f(p,q)} = \sum_{j=0}^{\min(m,n)} j!\, C_m^j C_n^j \left(\frac{h}{i}\right)^j \mathbf{q}^{n-j} \mathbf{p}^{m-j} \qquad (1)$$

where the bold face denotes the operators, e.g., $\mathbf{p} = -i\hbar\partial/\partial q$, and (1) was derived with the help of the Leibnitz formula and the commutation relation $[\mathbf{p,q}] = -i\hbar$.

Later, Moyal [5] treated the quantization problem as a problem of finding the phase-space distribution of a set of non-commuting operators **r** and **s** which in a specifically simple form are taken to be canonically- conjugate coordinates and momenta **q** and **p**. In particular, it was shown that an ordinary classical function G(p,q) is transformed into a quantum-mechanical operator-function according to the following prescription

$$\mathbf{G(p,q)} = \exp\left(\frac{h}{2i}\frac{\partial^2}{\partial p \partial q}\right) \mathbf{G_0(q,p)} \qquad (2)$$

where $\mathbf{G_0(q,p)}$ is obtained from G(p,q) by writing all the p's to the right of q's and then replacing p by the operator **p**. If we apply (2) to the classical expression $p^m q^n$ considered by Born and Jordan then we obtain the following expression

$$p^m q^n \to \mathbf{f}(\mathbf{p},\mathbf{q}) = \sum_{j=0}^{\min(m,n)} \left(\frac{h}{2i}\right)^j j!\, C_m^j C_n^j\, \mathbf{q}^{n-j}\, \mathbf{p}^{m-j} \qquad (3)$$

where each term differs from (1) by a factor $(1/2)^j$. More general quantization procedure was proposed in [9] which in a sense unified different quantization rules ( including Moyal's and Born-Jordan) by using the so-called s-parameterized ( $s \subseteq \mathbf{C}$) displacement operators

$$\mathbf{D}(s,\xi,\eta) = \exp(-is\frac{h\xi\eta}{2})\, e^{i(\xi q + \eta p)} \qquad (4)$$

Following Moyal we show that these operators would result in the quantization rule whose particular case $s=0$ yields the Moyal expression (2). Using Baker-Campbell-Hausdorf formula we rewrite (4) as follows

$$\mathbf{D}(\mathbf{p},\mathbf{q}, s,\xi,\eta) = \exp(i\frac{h\eta\xi(1-s)}{2})\exp(i\xi\mathbf{q})\exp(i h\mathbf{p}) =$$

$$\exp\left(\frac{h\eta(1-s)}{2}\frac{\partial}{\partial q}\right) e^{i\xi q} \exp\left(\frac{h\eta(1+s)}{2}\frac{\partial}{\partial q}\right) \qquad (5)$$

The characteristic function, or the expectation value, corresponding to the displacement operator $\mathbf{D}$ is

$$D(s, \xi, \eta) = \int \Psi^*\, \mathbf{D}(\mathbf{p},\mathbf{q}, s,\xi,\eta)\Psi\, dq \qquad (6)$$

Upon substitution of (5) in (6) we obtain

$$D(s, \xi, \eta) = \int \Psi^*\left(q - \frac{h\eta(1-s)}{2}\right) e^{i\xi q} \Psi\left(q - \frac{h\eta(1+s)}{2}\right) \qquad (7)$$

from which follows the Wigner formula [6] as a special case for s=0.

Introducing the momentum expansion

$$\Psi = \frac{1}{\sqrt{h}} \int \Phi(p)\, e^{ipq/h} dp$$

and using a new variable $Q = q - \frac{h\eta(1-s)}{2}$ we obtain from (7) that

$$D(s, \xi, \eta) = \frac{1}{\sqrt{h}} \int e^{i(\xi Q + \eta p)}\left\{\exp\left(\frac{h(1-s)}{2i}\frac{\partial^2}{\partial p \partial Q}\right)[\Psi^*(Q)\Phi(p)\, e^{ipQ/h}]\right\} dp dQ$$

is nothing more than the Fourier-transform of the phase-space distribution function F(s,p,q):

$$F(s,p,q) = \frac{1}{\sqrt{h}} \exp(\frac{h(1-s)}{2i} \frac{\partial^2}{\partial p \partial Q})[\Psi^*(Q)\Phi(p)\ e^{ipQ/h}] \qquad (8)$$

On the other hand, the average value of the classical function with the phase-space distribution function F(s,p,q), eq.(8) is:

$$<G(p,q)> = \int G(p,q)F(s,p,q)\ dpdq =$$

$$\frac{1}{\sqrt{h}} \int G(p,q) \exp(\frac{h(1-s)}{2i} \frac{\partial^2}{\partial p \partial q})[\Psi^*(q)\Phi(p)\ e^{ipQ/h}]\ dp\ dq \qquad (9)$$

Integrating (9) by parts we get

$$<G(p,q)> = \int \Psi^*(q)\ \{ \exp(\frac{h(1-s)}{2i} \frac{\partial^2}{\partial p \partial q})\ G_0(p,q)\}\ \Psi(q)\ dq$$

which means that the operator $\mathbf{G}(s,\mathbf{p},\mathbf{q})$ corresponding to the classical function G(p,q) is

$$\mathbf{G}(s,\mathbf{p},\mathbf{q}) = \exp(\frac{h(1-s)}{2i} \frac{\partial^2}{\partial p \partial q})\ G_0(p,q) \qquad (10)$$

where $G_0(p,q)$ is obtained from G(p,q) by writing all the operators $\mathbf{p}$ to the right of q. It is clear that the Moyal quantization rule (2) is a particular case of (10) with s = 0.

We can arrive at the quantization rule (10), or (2) by taking a different approach which is directly related with the correspondence $p \to \mathbf{p} = -ih\partial/\partial q$. Let us consider a classical function G(q,p) where we replace p by $p - ih\partial/\partial q \equiv p + \mathbf{p}$. It is clear that the resulting operator - function $\mathbf{G}(p - ih\partial/\partial q, q)$ depends on the ordering of the arguments. Generally speaking, we can represent $\mathbf{G}(p - ih\partial/\partial q, q)$ as follows

$$\mathbf{G}(p - ih\partial/\partial q, q) = G(\alpha \frac{h}{i}\frac{\partial}{\partial q} + p + (1-\alpha)\frac{h}{i}\frac{\partial}{\partial q}, q)$$

where $\partial/\partial q$ acts to the right of itself. The obtained expression can be rewritten as follows

$$\mathbf{G}(\mathbf{p},q) = \exp(\alpha \frac{h}{i} \frac{\partial}{\partial q} \frac{\overrightarrow{\partial}}{\partial p}) \, G(p,q)_{|p=0} \, \exp(\frac{(1-\alpha)h}{i} \frac{\overleftarrow{\partial}}{\partial p} \frac{\partial}{\partial q}) \qquad (11)$$

Here $G(p,q)_{|p=0}$ means that after differentiation we set $p = 0$, $\frac{\overrightarrow{\partial}}{\partial p}$ acts to the right and $\frac{\overleftarrow{\partial}}{\partial p}$ acts to the left. Using $\alpha = (1-s)/2$ we obtain from (11) the following expression

$$\mathbf{G}(\mathbf{p},q) = \sum_n \sum_{k=0}^{n} \sum_{t=0}^{k} \, (\frac{h}{2i})^n \, [C_{n-t}^{n-k}(1-s)^{k-t}(1+s)^{n+k}] \frac{(1-s)^t}{t!(n-t)!} [(\frac{\partial}{\partial p})^n G^{(t)}(p,q)]_{|p=0} (\frac{\partial}{\partial q})^{n-t}$$

where $G^{(t)}(p,q) \equiv (\partial/\partial q)^t G(p,q)$. Summation over k yields

$$\mathbf{G}(\mathbf{p},q) = \exp[\frac{h(1-s)}{2i} \frac{\partial^2}{\partial p \partial q}] \sum_m \frac{\partial^m G(p,q)}{\partial p^m}\bigg|_{p=0} \mathbf{p}^m \qquad (12)$$

The last sum is nothing more than the expansion of the operator-function $\mathbf{G}_0(\mathbf{p},q)$ in power series in $\mathbf{p}$ where all $\mathbf{p}$'s are placed to the right of q. Thus a rather straightforward procedure yields a generalization of the Moyal formula (2) for the case where $s \neq 0$. The derivation leading to (12) also proves the ordering in the Moyal quantization formula postulated without proof in the original paper by Moyal.

Here we would like to make two comments. First, the Moyal quantization is based upon a linear combination of the conjugate variables p and q. Therefore if we perform the canonical transformations from these variables to another set of conjugate variables P and Q then in general the original linear combination of p and q will not result in a linear combination of the new variables P and Q. Under such transformation the classical function G(q,p) goes to another function F(Q,P) = G( q(Q,P), p(Q,P)). This means that in this case we cannot apply the Moyal quantization formula expressed in terms of P and Q to the function of the new variables F(Q,P) despite of the fact that a canonically-conjugate transformation implies $\partial^2/\partial p \partial q = \partial^2/\partial P \partial Q$. An explicit example of such a case is given in the paper in [10].

Second, the derivation of the Moyal formula (2) is purely mathematical in a sense that it does not involve the dynamic laws of mechanics, i.e., the Hamilton equations. Therefore if we change from one set of canonically conjugate variables to another set then ( as we have indicated in the first comment) we cannot apply formula (2) without a certain modification. It is possible to demonstrate that the generalized quantization formula (10) supplemented by the physical condition allows one to correctly quantize canonically conjugate transformations.

Let us return to the Moyal formula (2) and notice that it has one remarkable feature. The power of the exponent is

$$\frac{h}{i}\partial^2/\partial p \partial q \tag{13}$$

On the other hand, dp dq represents an elementary area $dA_{pq}$ in the phase space. A change to another set of canonically conjugate variables

$$P = f(q,p), \quad Q = g(q,p)$$

maps the phase space p-q onto the space P-Q. This means that the elemental area $dS_{PQ}$ in the new space is related to the elemental area $dA_{pq}$ as follows

$$dA_{pq} = J\, dS_{PQ}$$

where J is the Jacobian of the transformation from p, q to P,Q. As a result (13) yields

$$\frac{h}{i}\frac{\partial}{\partial A_{pq}} = J^{-1}\frac{h}{i}\frac{\partial}{\partial S_{PQ}} \tag{14}$$

If the canonical transformation is such that J =1 then the Moyal quantization can be viewed as the unique method that retains the phase space area-preserving feature of classical canonical transformations. We still have to remember that the Moyal quantization in its simplest form (s=0) definitely works for linear canonical transformations but not for all the canonical transformations. Fortunately, in the very important case of the harmonic oscillator when we transform to action-angle variables the Moyal quantization provides us with the right answer.

To demonstrate this let us consider 1-D harmonic oscillator. In classical mechanics the its coordinate and momentum are expressed in terms of the action variable I and the angle variable Q by the following well-known formulae:

$$q = \sqrt{2\frac{I}{m\omega}}\sin Q, \quad p = \sqrt{2m\omega I}\cos Q \tag{15}$$

Since the transformations of q and p are now of the form $G(Q)F(I)$ where $[I,Q]_{PB} = 1$ the Moyal formula (2) provides us with a simple algorithm for expressing operators **q** and **p** in terms of the operators **I** and **Q**:

$$G(Q)F(I) \to G(I + \frac{h}{i}\partial/\partial Q) F(I) \qquad (16)$$

Applying (16) to (15) we get

$$\mathbf{q} = \sqrt{\frac{2}{m\omega}} \sin(Q + \frac{h}{2i}\frac{d}{dI}) I^{1/2} = \frac{1}{2i}\sqrt{\frac{2}{m\omega}} [e^{iQ}(\mathbf{I} + \frac{h}{2})^{1/2} - e^{-iQ}(\mathbf{I} - \frac{h}{2})^{1/2}]$$

(17)

$$\mathbf{p} = \frac{1}{2}\sqrt{2m\omega} [e^{iQ}(\mathbf{I} + \frac{h}{2})^{1/2} - e^{-iQ}(\mathbf{I} - \frac{h}{2})^{1/2} \qquad (18)$$

Operators **q** and **p** are linear combinations of the following operators

$$\mathbf{a}^+ = e^{iQ}(\mathbf{I} + \frac{h}{2})^{1/2}, \; \mathbf{a} = e^{-iQ}(\mathbf{I} - \frac{h}{2})^{-1/2}$$

which in turn are nothing more than shifting operators

$$\mathbf{a}^+ = i\sqrt{\frac{m\omega}{2}}\mathbf{q} + \sqrt{\frac{1}{2m\omega}}\mathbf{p}, \qquad \mathbf{a} = i\sqrt{\frac{m\omega}{2}}\mathbf{q} - \sqrt{\frac{1}{2m\omega}}\mathbf{p}$$

(18a)
.

To proceed further we have to express the hamiltonian $H = \frac{1}{2m}[p^2 + m^2\omega^2 q^2]$ in terms of **I** and **Q**. First we find $q^2$ and $p^2$ from Eqs. (17) and (18)

$$\mathbf{q}^2 = -\frac{1}{2m\omega}\{e^{iQ}(\mathbf{I} + \frac{h}{2})^{1/2} e^{iQ}(\mathbf{I} + \frac{h}{2})^{1/2} - e^{-iQ}(\mathbf{I} - \frac{h}{2})^{1/2} e^{iQ}(\mathbf{I} + \frac{h}{2})^{1/2} +$$

$$- e^{iQ}(\mathbf{I} + \frac{h}{2})^{1/2} e^{-iQ}(\mathbf{I} - \frac{h}{2})^{1/2} + e^{-iQ}(\mathbf{I} - \frac{h}{2})^{1/2} e^{-iQ}(\mathbf{I} - \frac{h}{2})-\} \qquad (19)$$

and analogously

$$\mathbf{p}^2 = \frac{m\omega}{2}\{e^{iQ}(\mathbf{I} + \frac{h}{2})^{1/2} e^{iQ}(\mathbf{I} + \frac{h}{2})^{1/2} + e^{-iQ}(\mathbf{I} - \frac{h}{2})^{1/2} e^{iQ}(\mathbf{I} + \frac{h}{2})^{1/2} +$$

$$+ e^{iQ} (I + \frac{h}{2})^{1/2} e^{-iQ} (I - \frac{h}{2})^{1/2} + e^{-iQ} (I - \frac{h}{2})^{1/2} e^{-iQ} (I - \frac{h}{2})^{1/2} \} \quad (20)$$

As a result, the Hamiltonian becomes

$$\mathbf{H} = \frac{\omega}{2} \{ e^{iQ} (I + \frac{h}{2})^{1/2} e^{-iQ} (I - \frac{h}{2})^{1/2} + e^{-iQ} (I - \frac{h}{2})^{1/2} e^{iQ} (I + \frac{h}{2})^{1/2} \}$$

(21)

We simplify Eq.(21) by finding the explicit expressions for $(I + \frac{h}{2})^{1/2} e^{-iQ}$ and $(I - \frac{h}{2})^{1/2} e^{iQ}$:

$$(I + \frac{h}{2})^{1/2} e^{-iQ} = e^{-iQ} (I - \frac{h}{2})^{1/2} \quad (22)$$

$$(I - \frac{h}{2})^{1/2} e^{iQ} = e^{iQ} (I + \frac{h}{2})^{1/2} \quad (23)$$

However expressions (22) and (23) provide the answers only with the accuracy to a sign! This is due to the fact that the square root of an operator $(I - \frac{h}{2})$ has two values corresponding to the two values of the square root of one. In that sense the Moyal algorithm generating an operator $\mathbf{F}(\mathbf{q}, \mathbf{p})$ from a classical function $F(q,p)$ gives only the principal value of the square root of one and <u>misses</u> all the other values of this square root. Therefore, in addition to (22), (23) we must have two additional expressions corresponding to the negative value of the square root of one, that is

$$(I + \frac{h}{2})^{1/2} e^{-iQ} = - e^{-iQ} (I - \frac{h}{2})^{1/2} \quad (24)$$

$$(I - \frac{h}{2})^{1/2} e^{iQ} = - e^{iQ} (I + \frac{h}{2})^{1/2} \quad (25)$$

It is clear that the application of $(I + \frac{h}{2})^{1/2}$ [ and respectively $(I - \frac{h}{2})^{1/2}$ ] to (22) [(23)] and (24) [(25)] yields the same result $(I + \frac{h}{2}) e^{-iQ} = e^{-iQ} (I - \frac{h}{2})$ [ and $(I - \frac{h}{2}) e^{iQ} = e^{iQ} (I + \frac{h}{2})$ ].

Thus taking into account both signs we write

$$(\mathbf{I} + \frac{h}{2})^{1/2} e^{-iQ} = \pm e^{-iQ} (\mathbf{I} - \frac{h}{2})^{1/2} \tag{26}$$

$$(\mathbf{I} - \frac{h}{2})^{1/2} e^{iQ} = \pm e^{iQ} (\mathbf{I} + \frac{h}{2})^{1/2} \tag{27}$$

Inserting (26) and (27) in the expression for the hamiltonian (22) and retaining only the positive signs we get the following

$$\mathbf{H} = \omega \mathbf{I} \tag{28}$$

$$\mathbf{H} = \frac{\omega h}{2} \tag{29}$$

Since from the point of view of dynamics (Hamilton's equations) a hamiltonian is defined only with the accuracy to an arbitrary constant we can combine (28) and (29) into one expression:

$$\mathbf{H} = \omega(\mathbf{I} + h/2) \tag{30}$$

Now the Schroedinger equation has an especially simple form

$$\omega(\mathbf{I} + h/2) \Psi = E\Psi \tag{31}$$

where $\mathbf{I} = (h/i)\partial/\partial Q$. If we seek the solution to (31) in the form of $\Psi = Ae^{i\alpha Q}$ and require that the wave function be periodic with the period that is integer- multiple of $2\pi$ ( the angle variable Q varies from 0 to $2\pi$) it is immediate that $\alpha = n$ ( an integer).

## 3. Moyal Quantization of Holographic Representations

As a generalization of the basic problem of quantizing a flat two dimensional phase space parameterized by p and q we now turn to the problem of quantizing the parameterization of a curved two dimensional surface. In the case of a Riemann surface the curved surface can be represented by a collection of flat sheets connected together along branch cuts. Thus in this case the problem of quantizing parameterizations of the surface would appear to be very similar to the Moyal problem of quantizing flat p,q phase space, except that now the phase space quantizations on each sheet must be matched along the branch cuts. One might guess that such a system could be quantized by introducing a set of p,q variables with an index j which represented which sheet one was on. The annihilation and creation operators **a** = **p**+i**q** and **a**$^+$ = **p**-i**q** introduced in section 2 (eq.18a) are now replaced with sets $\{\mathbf{a}_j\}$ and $\{\mathbf{a}^+_j\}$. The original Weyl- Heisenberg group will be replaced by a Lie group generated by operators of the form $\alpha^* \mathbf{a} \equiv \sum_{j=1}^{N} \alpha^*_j \mathbf{a}_j$ and $\alpha \mathbf{a}^+ \equiv \sum_{j=1}^{N} \alpha_j \mathbf{a}^+_j$. States playing much the same role as the Glauber coherent states introduced in section 2 will be generated by the operators

$$\mathbf{D}(\alpha) = \exp(\alpha \mathbf{a}^+ - \alpha^* \mathbf{a}). \tag{32}$$

which are analogous to the displacement operators $\mathbf{D}(\xi,\eta)$. These operators obey the multiplication rule

$$\mathbf{D}(\alpha)\mathbf{D}(\beta) = e^{i\mathrm{Im}(\alpha\beta^*)} \mathbf{D}(\alpha+\beta). \tag{33}$$

Replacing the Weyl- Heisenberg algebra with an arbitrary Lie algebra leads to the generalized coherent states of Gilmore[11] and Perelomov [12]. For these generalized coherent states the space parameterized by $\alpha$ and $\alpha^*$ is no longer the complex plane but a symmetric space G/H. However these symmetric spaces have a natural symplectic structure [13] and provide a natural phase space structure for quantum systems whose dynamics is described by Lax Pair type equations. For example, in the SU(N) case in terms of the variables

$$z = \alpha \frac{\sinh\alpha\alpha^*}{\sqrt{\alpha\alpha^*}}$$

the metric of the space parametrized by the $\alpha$ variables is proportional to dzdz*, and the Poisson bracket takes the simple form

$$\{f,g\} = -i\sum_j \left[ \frac{\partial f}{\partial z_j^*}\frac{\partial g}{\partial z_j} - \frac{\partial g}{\partial z_j^*}\frac{\partial f}{\partial z_j} \right]. \tag{34}$$

Evidently the variables z and z* play essentially the same role as the p and q variables in ordinary mechanics, and therefore we expect that the formalism introduced in section 2 can be used to define operators on the 2N-dimensional phase space parameterized by α and α* and introduce a Wigner-like distribution function on this space. This circumstance suggests using Moyal quantization to quantize non-linear integrable models. In fact this line of investigation has been pursued in a number of recent papers [ see e.g. refs. 3 and 4].

It is interesting to note that when G/H is a complex torus then the coherent states generated by the **D**(α) can be identified with the theta functions that play such an important role in the theory of Jacobian varieties. Indeed the condition that the generalized coherent states be single valued on the torus means that the phase factor Im(αβ*) in equation (33) must be equal to 2π times an integer when α and β correspond to periods of the torus.Remarkably this is just the condition that the complex N-torus be an abelian variety; i.e. the Jacobian of a Riemann surface. Thus it is possible to regard the Jacobian of a Riemann surface as a kind of phase space and to use theta functions to define a quantization of this space a la Moyal-Wigner. Since according to the classical Torelli theorem a Riemann surface can be reconstructed from its Jacobian and associated theta functions, one might assume that applying the Moyal formalism to the Jacobian of the Riemann surface effectively solves the problem of quantizing the geometry of a Riemann surface. We will now argue that what is involved here is essentially the assumption that quantizing the Riemann surface is equivalent to quantizing holographic representations of the surface.

As a quick reminder all information concerning an arbitrarily curved surface in 3-dimensions can be encoded onto a flat plane or the surface of a sphere by recording photographically or otherwise the interference of light scattered off the surface with a reference beam. Of course, if the surface is reentrant then the interference pattern must be recorded for various orientations of the illuminating beam in order to capture the entire surface. In order to physically describe a hologram we must introduce a slowly varying "envelope" electric field whose rms magnitude corresponds to the intensity of the interference pattern. To this end we write the vector potential on the recording surface as a function of position x on the hologram in the form

$$\mathbf{A}(x) = -(i/k)\,\mathbf{E}(x,t)\,\exp(ikz) + \text{h.c.}, \qquad (35)$$

where z is a coordinate for the direction perpendicular to the surface of the hologram and the factor $\exp(ikz)$ represents the rapidly varying phase of the reference and scattered beams. It is straightforward to quantize these fields by substituting expression (35) into the standard radiation gauge Lagrangian for the electromagnetic field. One finds the following commutation relations for the electric field operators on the recording surface:

$$[\mathbf{E}_i(x),\,\mathbf{E}^+_j(x')] = 2\pi\omega\delta_{ij}\,\delta(x-x'). \qquad (36)$$

*We are here representing the geometry of the Riemann surface using ordinary harmonic oscillator creation and annihilation operators. However, as we showed in section 2 Moyal quantization involves using linear combinations of these operators. Therefore Moyal quantization of a Riemann surface evidently corresponds to quantum holography using squeezed photon states instead of the usual coherent states. The detailed nature of holographic representations with squeezed states is obviously a subject of interest in itself; however, our main purpose in this paper is to make contact with the classical theory of Riemann surfaces and associated integrable models.*

The electric field $\mathbf{E}(x,t)$ receives contributions from various points on the surface. In the dipole approximation the contribution from each little patch of surface is determined by the cross product between the vector pointing from the patch of surface to the point x and the direction of the oscillating polarization induced in the surface patch by the illuminating beam. Since these two vectors are curl-free on the surface the sum of contributions over the surface has the character of an inner product of two harmonic differentials:

$$(\omega_1,\,\omega_2) = \int_S \omega_1 \wedge \omega_2^* \qquad (37)$$

Now an inner product for harmonic differentials of the form (37) plays an important role in the theory of Riemann surfaces [7], and so here we are beginning to see the promised connection between the Torelli theorem and holography.

In one obviously important respect though the holographic representation using the electric field $\mathbf{E}(x,t)$ differs from the representation discussed earlier that involved N-copies the Weyl-Heisenberg group; namely, the holographic representation involves an infinite number of annihilation and creation operators corresponding to an infinite number of positions on the hologram, whereas the representation based on the product of Weyl-Heisenberg groups involved only a finite number of such operators. This discrepancy can be traced to the fact that the quantization problem most closely related to ordinary p,q phase space quantization assumes that the shape of the surface is fixed. Under such circumstances the electric field vectors at different x points are not independent for a given orientation of the illuminating beam. Indeed upon reflection it is clear that the only way to obtain algebraically independent annihilation and creation operators is to illuminate the different handles of the Riemann surface in distinct ways. An aesthetic choice for such distinct illuminations would be to choose the different illuminating beams in such a way that the polarizations induced on the surface by the different beams can be identified with a canonical basis for the first cohomology group. There are 2g 1-forms in such a canonical basis; one for each of the 2g 1-cycles associated with the g handles of the surface [7]. Armed with such canonical illuminations we can record a set of 2g algebraically independent holograms, from which the Riemann surface can be completely reconstructed. Corresponding to these 2g holograms are 2g independent operators, say $\mathbf{E}(x_0)$ for each hologram, thus confirming our previous guess concerning the structure of the quantized phase space for a Riemann surface.

Of course, in these considerations we have failed to exploit the infinite number of degrees of freedom associated with the $\mathbf{E}(x)$ operators. Taking into account the quantum fluctuations associated with these degrees of freedom would lead to what might legitimately called quantum holography. Such a theory might also be interpreted as a quantum theory of two dimensional geometry, where the classical geometry of the Riemann surface *in some sense* has been smeared out, and replaced with non-commuting operators.

## 4. Conclusion

We have seen that the quantization formalism of Moyal and Wigner can provide a natural way to quantize problems where the basic setting is a symplectic space. In the case of a Riemann surface one can view the symplectic structure as being provided either by the complex structure of the manifold or by the two types of non-trivial 1-cycles associated with the handles on the surface. These symplectic spaces can also be represented holographically. Quantization of the holographic representation of the complex structure leads to an infinite dimensional Hilbert space that resembles the Fock space for a free field theory. Remarkably this representation of a surface with fixed shape in terms of free field operators seems to be related to the well known fact that non-linear integrable models can often be transformed into linear field equations. For example, Plebanski's non-linear equation for self-dual Einstein spaces can be transformed into the Laplace equation by a Legendre transformation. In the previous section we were able to make direct contact with the classical Lax pair equations for integrable models by applying Moyal quantization to the symplectic space associated with the non-trivial 1-cycles of a Riemann surface. It is our hope that this formalism will prove useful in elucidating the physical meaning of quantized non-linear models. In particular, our holographic interpretation of quantized non-linear models may aid in the development of a quantum theory of membranes.

Hyperelliptic Riemann surfaces have a well known [14] association with $SU(N)$ Lie algebras and Toda models. Furthermore the $N \to \infty$ limit of this association is apparently closely related to a theory of membranes. For example, it is believed that the sought after quantum theory of membranes can be represented as a supersymmetric $SU(\infty)$ Yang-Mills theory that depends on one temporal dimension; i.e. $SU(\infty)$ quantum mechanics [15, 16]. Now this $SU(\infty)$ quantum mechanics contains a 2-dimensional $SU(\infty)$ Toda model [17] and it has been shown [18, 19] that this Toda model can be interpreted as a theory of self-dual membranes. The Moyal quantization of this Toda theory was studied in reference 4. More recently Fairlie [20] has applied Moyal quantization directly to $SU(\infty)$ quantum mechanics. Unfortunately the physical interpretation of these quantizations is unclear. Moyal quantization of holographic representations may provide the needed clue. For example, the reduction of a general theory of membranes to an integrable theory of self-dual membranes is reminiscent of our reduction from a general quantum theory of holography of curved surfaces to the simpler theory based on quantization of a Jacobian variety.

Acknowledgement. We are very grateful to Carlos Castro for stimulating conversations and encouragement during the writing of this paper.

----------------------------------------------------------------

+ Lawrence Livermore National Laboratory., Livermore, CA  94550

++ Physics Dept., Univ.of  the Pacific, Stockton, CA  95211